\documentclass[12pt]{article}
\usepackage{amsmath}
\usepackage{setspace}
\usepackage{graphicx,psfrag,epsf}
\usepackage{natbib}
\usepackage{amssymb,amsfonts,amsthm,amsmath,color}
\usepackage{multirow}
\usepackage[toc,page]{appendix}
\usepackage{mathrsfs}
\usepackage{enumitem}
\usepackage{authblk}

\usepackage{algorithm}
\usepackage{algpseudocode}
\usepackage{url} 
\usepackage{graphicx} 
\usepackage[justification=centering]{caption}
\usepackage{subfig} 
\newtheorem{proposition}{Proposition}[section]

\newcommand{\bm}[1]{\mbox{\boldmath$ #1 $\unboldmath}}

{
        \title{\bf D-optimal Design for Network A/B Testing}
        \date{}
        \author[1]{Victoria Pokhiko}
        \author[2]{Qiong Zhang}
        \author[3]{, Lulu Kang\thanks{Corresponding author; Email: lkang2@iit.edu}}
        \author[1]{, D'arcy P. Mays}
\affil[1]{Department of Statistical Sciences and Operations Research, Virginia Commonwealth University}
\affil[2]{School of Mathematical and Statistical Sciences, Clemson University}
\affil[3]{Department of Applied Mathematics, Illinois Institute of Technology}
    } 

\begin{document}
\maketitle
\linespread{1.0}
\begin{abstract}
A/B testing refers to the statistical procedure of conducting an experiment to compare two treatments, A and B, applied to different testing subjects. 
It is widely used by technology companies such as Facebook, LinkedIn, and Netflix, to compare different algorithms, web-designs, and other online products and services. 
The subjects participating these online A/B testing experiments are users who are connected in different scales of social networks.
Two connected subjects are similar in terms of their social behaviors, education and financial background, and other demographic aspects. 
Hence, it is only natural to assume that their reactions to the online products and services are related to their network adjacency. 
In this paper, we propose to use the conditional auto-regressive model to present the network structure and include the network effects in the estimation and inference of the treatment effect. 
A D-optimal design criterion is developed based on the proposed model. 
Mixed integer programming formulations are developed to obtain the D-optimal designs. 
The effectiveness of the proposed method is shown through numerical results with synthetic networks and real social networks. 
\end{abstract}
\newpage

\section{Introduction}

The theory of A/B testing dates back to Ronald Fisher's experiments at the Rothamsted Agricultural Experimental Station in England in the 1920s \citep{yates1964sir}. 
A standard statistical testing framework is the Rubin causal model \citep{rubin1974estimating} usually used to conduct and analyze A/B testing experiments.
A key assumption made in the Rubin causal model is the Stable Unit Treatment Value Assumption (SUTVA), which states that the behavior of each test subject in the experiment depends only on the individual treatment and not on the treatments of others, i.e., the test subjects are independent. 

Recently, A/B testing has been widely used online to test which alternative or treatments out of the two, A or B, leads to better outcomes. 
The treatments could be options of online commercial, web page designs, different recommendation algorithms, or any new online features that need to be evaluated so that the companies can make informed decisions. 
The response measures of the experiments can be numerical values of profits, sales, return on investment, click through rate, etc. 
Usually, the participants of the experiments are sampled from a much bigger population of users. 
Then the experimenter randomly assigns those subjects to either the treatment or control group. 
This procedure works well when the subjects can be considered independent of each other.  

However, in a social network environment, a user is more likely to adopt a new product or service if people around him/her adopt it too. 
An individual's behavior can have a non-trivial effect on his/her social network.
This effect is called network effect, also known as social interactions, peer influence, or social interference \citep{aronow2012estimating, eckles2017design, athreya2017statistical}. 
In an A/B testing experiment, this implies that if the treatment has a significant impact on a subject, the effect would reach his/her social circles, regardless whether his/her neighbors are in the treatment or control group. 
To account for network connections in causal analyses, researchers usually work with two specific settings, network interference, and network-correlated outcomes. 
When the network interference is present, the outcome of node $i$ (or user $i$) is a function of the treatment assigned to node $i$ and the treatment assigned to other nodes that are related to node $i$ through the network, and possibly the observed outcomes of these related nodes. 
For the network-correlated outcomes, the outcomes of the neighboring nodes are correlated because the features of the two connected nodes are more similar than those of the unconnected nodes \citep{basse2017limitations, basse2018model}.
This paper focuses on network-correlated outcomes. 
Under this setting, we assume that the A/B testing outcomes of adjacent nodes are positively correlated due to their similarities.

An important question in A/B testing is how to allocate treatments to the subjects. 
Different from the SUTVA assumption, a random design which randomly assigns the treatment settings with equal probability to each user may not be efficient in estimating the treatment effect in the presence of the network-correlated outcomes. 
Cluster-based randomized treatment allocation has been used to block the effect of network correlation in A/B testing experiments. 
One such example can be found in \cite{xu2015infrastructure}. 
Also, \cite{saveski2017detecting} and \cite{pouget2017testing} used the cluster-based random design to determine the existence of the network effect. 
\cite{basse2018model} proposed the restricted randomization approach to minimize the mean squared error of the estimated treatment effect. 
Based on a normal-sum model, the analytical decomposition of the mean square error provided insights to develop the restricted randomization strategies in the absence of a detailed network structure. 
Although these cluster-based random designs are simple to use, they might not be able to achieve a perfect balance between the two treatment groups in terms of their network structures. 
If a reasonable model can be assumed for the effects of the treatment and network \citep{chen2018sequential}, the classic model-based optimal design \citep{atkinson2007optimum, wu2011experiments} can also be used for A/B testing experiments. 
Unfortunately, there has not been much development in this direction and we decide to fill the gap. 

In this paper, we focus on the construction of A/B testing experimental designs for network-correlated outcomes when users who are connected in a network share some common social and demographic backgrounds. 
We propose a spatial network model for A/B testing, called conditional auto-regressive model or CAR \citep{schmidt2014conditional} to incorporate the correlated network structure in the analysis. 
To accurately estimate the treatment effect, we use the D-optimal criterion \citep{sitter1995d}, which seeks to maximize the determinant of the information matrix of the linear regression model of the response with respect to the treatment effects and other potential variables.
Mixed integer programming formulations are developed to optimize the D-optimal criterion and construct the design. 
Finally, we conduct simulation studies on synthetic and real social networks to demonstrate the performances of the proposed method compared to the random designs, which do not consider the network structure.

\section{D-Optimal Design for CAR Model} 
\subsection{Network A/B Testing with CAR Model}

We consider an A/B testing experiment conducted on a social network with $n$ nodes. 
The social network is considered to be an undirected graph in the context of this paper. 
The edges of this network are recorded by an $n\times n$ adjacent matrix $W$ whose $(i,j)$-th entry is $w_{ij}$. 
The diagonal entries $w_{ii}$'s of this matrix is 0, whereas off-diagonal entries are
\begin{equation}
w_{ij}=\begin{cases}
1, & \text{if node $i$ and node $j$ are adjacent}\\
0, & \text{otherwise}.
\end{cases}
\end{equation} 
Two adjacent nodes are the ones connected by an edge. 
The experimental design is the plan to allocate A or B treatment to each node. 
Let $x_i\in \{1,-1\}$ for $i=1, \ldots, n$ be the design of the $i$-th node and the two settings $\{1,-1\}$ represent A and B treatments. 
Denote the scalar response observation of the $i$-th node by $y_i$. 
In this paper, we focus on the case that the response is continuous. 
Assume a linear regression model for the response as follows. 
\begin{equation}\label{eq:gcar}
y_i=\beta_0+x_i\beta+\delta_i,
\end{equation}
where $\beta_0$ is the intercept, $\beta$ represents the treatment effect, and $\delta_i$ is a zero mean random variable. 
Under the SUTVA assumption, $\delta_i$'s are assumed to be the random noise and independent with each other. 
But for the experiments on networks, two connected users share similarities in their social behaviors and other backgrounds, and thus their responses are often correlated. 
To incorporate this social correlation, we model $\delta_i$ in \eqref{eq:gcar} by the conditional auto-regressive (CAR) model \citep{besag1974spatial}
\begin{equation}\label{eq:delta}
\delta_i|\boldsymbol{\delta}_{-i}\sim N\left(\rho\sum_{j\neq i}\frac{w_{ij}\delta_j}{m_i}, \frac{\sigma^2}{m_i}\right),
\end{equation}
where $\boldsymbol{\delta}_{-i}=\{\delta_1, \ldots, \delta_{i-1}, \delta_{i+1}, \ldots, \delta_n\}$, $m_i=\sum^n_{j=1}w_{ij}$ is the number of nodes adjacent to the $i$-th node, $\sigma^2$ is the variance parameter that is assumed to be a constant in our scope, and $|\rho|<1$ is the correlation parameter of the CAR model. 
If $\rho=0$, $\delta_i$'s are independent with each other, which corresponds to the extreme case when the network only has $n$ nodes but without any edges. 
As noted in the Introduction, the connected users tend to have similar reactions to the same treatment. 
Hence, without loss of generality, we restrict that the correlation parameter is non-negative, i.e., $0\leq\rho<1$.

By the Brook's Lemma \citep{brook1964distinction}, $\boldsymbol{\delta}=(\delta_1, \ldots, \delta_{n})^\top$ follows a multivariate normal distribution:
\begin{equation}\label{eq:mvn}
    \boldsymbol{\delta}\sim \mathcal{MVN}_n(0, \sigma^2(D-\rho W)^{-1})
    \end{equation}
where $D=\mathrm{diag}(m_1, \ldots, m_n)$. The detailed 
derivation to \eqref{eq:mvn} is deferred to Appendix \ref{sec:A1}.
The maximum likelihood method can be used to fit the model \eqref{eq:gcar} with $\delta_i$ from \eqref{eq:delta} and estimate the model parameters. 
The goal of A/B testing is to accurately assess the treatment effect $\beta$. 
Next, we determine values of $x_i$'s using D-optimal design to improve the accuracy of the estimate $\hat\beta$.  

\subsection{D-optimal Design for CAR Model }

Given the correlation parameter $\rho$, the parameters $\beta_0$ and $\beta$ in \eqref{eq:gcar} can be estimated by 
\begin{equation}
    (\hat\beta_0, \hat\beta)^\top=(X^\top V^{-1} X)^{-1} X V^{-1} \bm y,
\end{equation}
where $X$ is a $n\times 2$ design matrix with $i$-th row $(1, x_i)$, $\bm y=(y_1, \ldots, y_n)^\top$, and $V=(D-\rho W)^{-1}$, and the variance matrix of 
$(\hat\beta_0, \hat\beta)^\top$ is 
\begin{equation}\label{eq:varm}
    \mathrm{Var}\left\{(\hat\beta_0, \hat\beta)^\top\right\}=\sigma^2(X^\top V^{-1}X)^{-1}=\sigma^2(X^\top (D-\rho W)X)^{-1}.
\end{equation}

Under our model assumption in \eqref{eq:gcar}, the D-optimal design is determined by maximizing the determinant of the matrix $X^\top (D-\rho W)X$ in \eqref{eq:varm} with respect to $\bm x=(x_1, \ldots, x_n)^\top\in \{-1, 1\}^n$, i.e.,
\begin{equation}\label{eq:dopt}
    \mathrm{argmax}_{\bm x\in \{-1,1\}^n} \{D(\bm x):=|X^\top (D-\rho W)X|\}.
\end{equation}
Notice that,
\[\left\{X^\top (D-\rho W)X\right\}^{-1}  = \begin{bmatrix}
        \sum\limits_i m_i-\rho \sum\limits_i \sum\limits_j w_{ij} & \sum\limits_i m_i x_i-\rho \sum\limits_i \sum\limits_{j>i}  w_{ij} (x_i + x_j)\\
        \sum\limits_i m_i x_i-\rho \sum\limits_i \sum\limits_{j>i}  w_{ij} (x_i + x_j) & \sum\limits_i m_i x_i^2-\rho \sum\limits_i \sum\limits_j  w_{ij} x_i x_j
        \end{bmatrix}^{-1}    
\]        
\[ = \frac{1}{D(\bm x)} \begin{bmatrix}
        \sum\limits_i m_i x_i^2-\rho \sum\limits_i \sum\limits_j w_{ij} x_i x_j
         & -(1-\rho)\sum\limits_i m_i x_i\\
        -(1-\rho)\sum\limits_i m_i x_i & (1-\rho)\sum\limits_i m_i
        \end{bmatrix}.
    \]
Then we obtain the following
\begin{equation}\label{eq:dx}
D(\bm x) = (1-\rho)\sum\limits_i m_i (\sum\limits_i m_i -\rho \sum\limits_i \sum\limits_j  w_{ij} x_i x_j) - (1-\rho)^2(\sum\limits_i m_i x_i)^2,
\end{equation}
and
\[
\mathrm{Var}(\hat{\beta})=\frac{\sigma^2(1-\rho)\sum\limits_i m_i}{D(\bm x)}. 
\]
Given a network, $\sigma^2(1-\rho)\sum\limits_i m_i$ is a constant. Therefore, the D-optimal design also minimizes the variance of the treatment effect.  
Based on \eqref{eq:dx}, Proposition \ref{prop:dopt} gives a simplified objective function to obtain the D-optimal design. 

\begin{proposition}\label{prop:dopt}
Under model assumptions \eqref{eq:gcar} and \eqref{eq:delta}, the D-optimal design $\bm x$ is the solution of the following optimization problem. 
\begin{equation}\label{eq:obj}
     \mathrm{argmax}_{\bm x\in \{-1,1\}^n} D(\bm x) = \mathrm{argmin}_{\bm x\in \{-1,1\}^n} \left\{a \sum\limits_i \sum\limits_j w_{ij} x_i x_j + \left(\sum\limits_i m_i x_i\right)^2\right\},
\end{equation}
where $a=\frac{\rho}{1-\rho}\sum^n_{i=1} m_i$.
\end{proposition}
The proof of this proposition is provided in the Appendix A2.
Since $a$ is non-negative, a lower bound of the objective function in \eqref{eq:obj} can be attained by minimizing $\sum\limits_i \sum\limits_j w_{ij} x_i x_j$ and $\left(\sum\limits_i m_i x_i\right)^2$, respectively. 
The lower bound of $\sum\limits_i \sum\limits_j w_{ij} x_i x_j$ is $-\sum\limits_i\sum\limits_j w_{ij}$ if 
$x_i\neq x_j$ for all $(i,j)$ with $w_{ij}=1$.  The lower bound of $\left(\sum\limits_i m_i x_i\right)^2$
is zero if $\sum\limits_i m_i x_i=0$, which represents that 
the nodes allocated with -1 and the nodes allocated with 1 have equal number of first order neighborhoods. Similarly, we can obtain an upper bound for the D-optimality measure: 
\begin{equation}
  D(\bm x)\leq (1-\rho)(\sum^n_{i=1}m_i)^2+(1-\rho)\rho\sum^n_{i=1}m_i\sum^n_{i=1}\sum^n_{j=1}w_{ij}.
\end{equation} 
According to this upper bound, we are able to define a D-efficiency measure:
\begin{equation}\label{eq:d-eff}
\frac{D(\bm x)}{(1-\rho)(\sum^n_{i=1}m_i)^2+(1-\rho)\rho\sum^n_{i=1}m_i\sum^n_{i=1}\sum^n_{j=1}w_{ij}}.
\end{equation}
This D-efficiency measure ranges from 0 to 1, which evaluates the quality of the design without concerning the scale of the $D(\bm x)$. 
A larger value of D-efficiency corresponds to a better design. 
The definition of the D-efficiency is different from the conventional version in literature \citep{atkinson2007optimum}, which would be the $1/p$th root of \eqref{eq:d-eff} for $p$ experimental factors. 
But there is no need to calculate the $1/p$th root because only one experimental factor and no other covariates are involved in the CAR model. 

\section{Mixed Integer Programming Formulations for D-optimal Design}\label{sec3}

Since the decision space in \eqref{eq:dopt} is $\{-1, 1\}^n$, the optimization problem is an integer programming problem \citep{nemhauser1988integer}. 
To solve this problem, this section formulates the D-optimal design problem into mixed integer programming problems with the original D-optimal objective function in \eqref{eq:obj} and
a modified D-optimal objective function. 

\subsection{A Mixed Integer Programming Formulation for D-optimal Design}\label{sec:mip}  
By observing that $(\sum\limits_i m_i x_i)^2 = \sum\limits_i m_i^2 x_i^2 + \sum\limits_i \sum\limits_{j \ne i} m_i m_j x_i x_j = \sum\limits_i m_i^2 + \sum\limits_i \sum\limits_{j \ne i} m_i m_j x_i x_j$, we express the objective function in \eqref{eq:dopt} by
\[a \sum\limits_i \sum\limits_j w_{ij} x_i x_j + \sum\limits_i \sum\limits_{j \ne i} m_i m_j x_i x_j =
\sum\limits_i \sum\limits_{j \ne i} b_{ij} x_i x_j,
    \]
where $b_{ij} =  a w_{ij} + m_i m_j $. 
By introducing new variables $v_i=(x_i+1)/2$ and $u_{ij}=v_iv_j$, we can formulate the original optimization problem \eqref{eq:obj} into the following mixed integer linear program (MIP) problem \eqref{eq:omip}. 
Although $u_{ij}$ is not restricted to be binary, its definition makes $u_{ij}$ can only be 0 or 1. 
\begin{align}
\label{eq:omip}
\mathrm {min} & \left[\sum\limits_i \sum\limits_{j \neq i}  b_{ij} u_{ij} - \sum\limits_i \sum\limits_{j \neq i}  b_{ij} v_i \right] \\\nonumber
\textrm{subject to }& u_{ij} \leq v_i, \quad u_{ij} \leq v_i\\\nonumber
& u_{ij} \geq v_i+v_j-1, \\\nonumber
& u_{ij} \in \mathbb{R}, \quad u_{ij} \geq 0,\\\nonumber
& v_i \in \{0, 1\}, \textrm{ for }i=1,\ldots, n, j=1, \ldots, n. 
\end{align}

\begin{figure}[htb]
\centering
\includegraphics[scale=0.3]{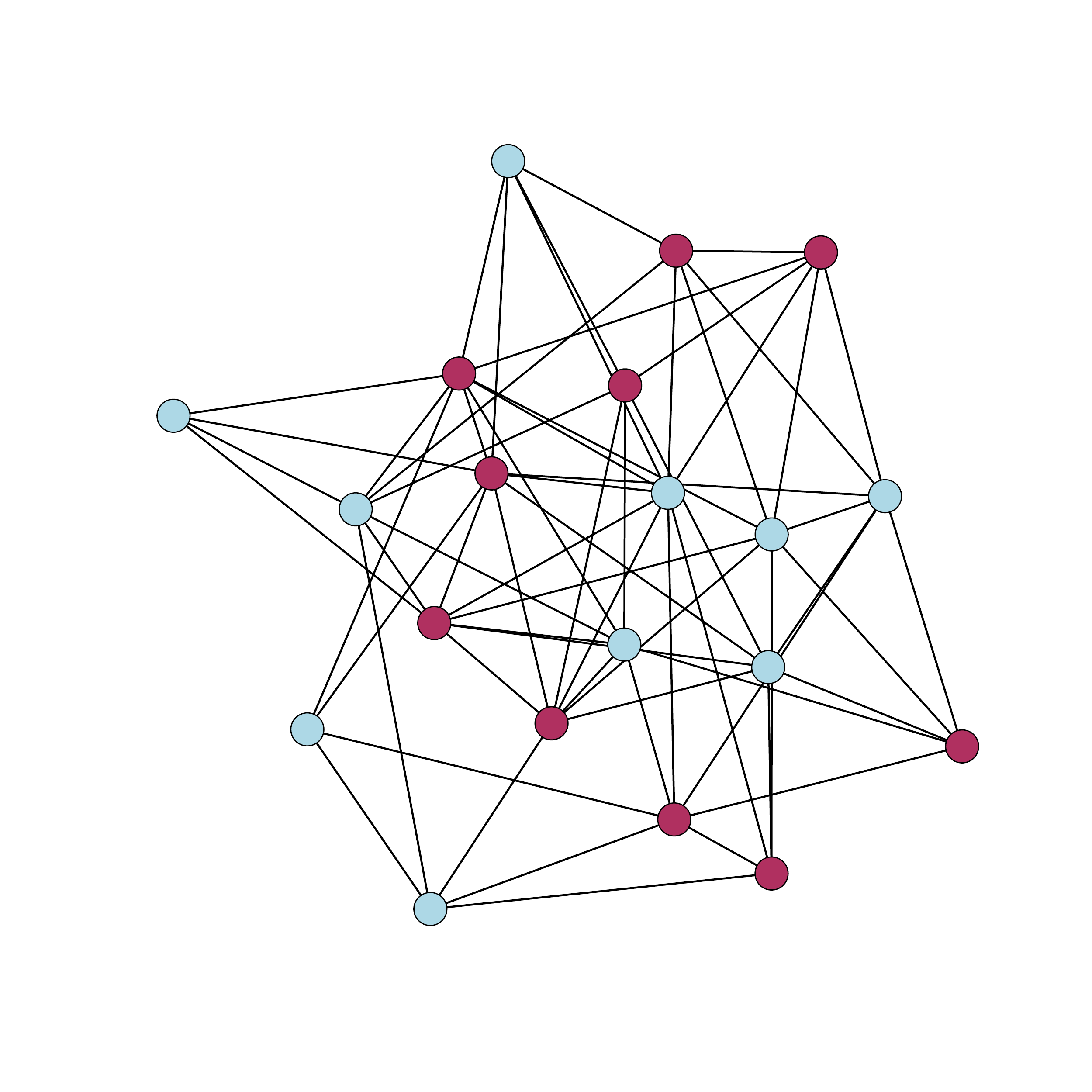}
\captionof{figure}{D-optimal Design obtained from the mixed integer programming formulation in Section \ref{sec:mip}. Two colors represents -1 or 1 allocation.}
\label{fig:plot}
\end{figure}
Figure \ref{fig:plot} depicts an example of solving this MIP problem to construct design to a network of size 20.
Notice that $b_{ij}$ depends on the value of $a$, which is a function of the correlation parameter $\rho$ in \eqref{eq:delta}. 
It is impractical to assume that $\rho$ can be accurately estimated before data collection. 
Using the Bayesian framework, we can derive the Bayesian D-optimal design criterion which is the expectation of the D-optimality $D(\bm x)$ with respect to a user-specified prior distribution of $\rho$. 
From \eqref{eq:obj}, we can see that the expected D-optimality only depends on the expected value of the ratio $\rho/(1-\rho)$, which essentially is a tuning parameter. 
Instead of the Bayesian approach, we decide to take an equivalent route by modifying the objective function and moving this tuning parameter into the constraint. 

\subsection{A Mixed Integer Programming Formulation for a Modified D-optimality Criterion}\label{sec:modified-mip}

As we point out above, reducing values of $\sum\limits_i \sum\limits_j w_{ij} x_i x_j$ or $(\sum\limits_i m_i x_i)^2$ would improve  D-efficiency defined in \eqref{eq:d-eff}.  
To remove the parameter $a$ from the objective function in \eqref{eq:dopt}, an alternative solution is to use $\sum\limits_i \sum\limits_j w_{ij} x_i x_j$ as the objective function, and bound the value of $(\sum\limits_i m_i x_i)^2$ by a constraint. 
We modify the optimization problem in \eqref{eq:dopt}
to be 
\begin{equation} \label{prob:mod1}
 \text{min}_{\bm x\in \{-1, 1\}^n} \sum\limits_i \sum\limits_j w_{ij} x_i x_j 
\end{equation}
\[\text{s.t. } 
 -\delta \leq\sum\limits_i m_i x_i \leq \delta \text{, and } x_i\in \{-1,1\}~\mathrm{for}~i=1, \ldots, n, 
\]
where $\delta>0$ is a tuning parameter. 

 For different networks, the ranges of $\sum\limits_i m_i x_i$ can be different. 
Now we discuss how to specify the value of $\delta$ by normalizing it to a unified range for different networks. 
Assume that $x_i$ is a random variable taking value from $\{-1, 1\}$ with equal weights. Hence, 
\[
\mathrm{E}\left(\sum^n_{i=1} m_i x_i\right)=\sum^n_{i=1} m_i \mathrm{E} x_i=0 ~\mathrm{and}~
\mathrm{Var}\left(\sum^n_{i=1} m_i x_i\right)=\sum^n_{i=1} m^2_i \mathrm{Var}(x_i)=\sum^n_{i=1} m^2_i.
\]
If $n^{-2}\sum^n_{i=1} m^2_i<\infty$ and $n^{-1}m_i\rightarrow 0$ for $i=1,\ldots, n$,
we have that 
$\sum^n_{i=1} m_i x_i/\sqrt{\sum^n_{i=1} m^2_i}$ asymptotically follows a standard normal distribution by the Lindeberg's Central Limit Theorem.
Therefore, as $n\rightarrow\infty$, 
\[
\mathrm{P}\left(-\Phi^{-1}(\alpha)\sqrt{\sum\limits_i m^2_i}<\sum m_i x_i<\Phi^{-1}(\alpha)\sqrt{\sum\limits_i m^2_i}\right)=2\alpha-1,
\]
where $\Phi(\cdot)$ is the cumulative distribution function of the standard normal distribution, and $\alpha\in(0.5, 1)$. 
By
\[
\delta =\Phi^{-1}(\alpha)\sqrt{\sum\limits_i m^2_i},
\]
$\delta$ is increasing with $\alpha$. 

Define $v_i$ and $u_i$ for $i=1,\ldots,n$ in the same way as in Section \ref{sec:mip}, the problem in \eqref{prob:mod1} becomes
\begin{align}
\label{prob:mod2}
\text{min } & \left[ \sum\limits_i \sum\limits_{j \neq i}  w_{ij} u_{ij} - \sum\limits_i \sum\limits_{j \neq i}  w_{ij} v_i\right] \\\nonumber
\textrm{subject to }
& \sum\limits_i m_i v_i \leq \frac{1}{2}\left(\sum\limits_i m_i  + \delta\right) \\\nonumber
& \sum\limits_i m_i v_i \geq \frac{1}{2}\left(\sum\limits_i m_i  - \delta\right) \\\nonumber
& u_{ij} \leq v_i, \quad u_{ij} \leq v_i\\\nonumber
& u_{ij} \geq v_i+v_j-1, \\\nonumber
& u_{ij} \in \mathbb{R}, \quad u_{ij} \geq 0,\\\nonumber
& v_i \in \{0, 1\}, \textrm{ for }i=1,\ldots, n, j=1, \ldots, n. 
\end{align}

Here are some remarks on the two reformulated optimization problems \eqref{eq:omip} and \eqref{prob:mod2}. 
First, solving the MIP \eqref{eq:omip} in Section \ref{sec:mip} gives the exact D-optimal design, whereas solving the MIP \eqref{prob:mod2} for the modified problem in Section \ref{sec:modified-mip} does not guarantee that the exact D-optimal design can be found. 
Second, the MIP \eqref{eq:omip} requires that the correlation parameter $\rho$ to be known, whereas the modified MIP \eqref{prob:mod2} does not. 
Third, the original numbers of decision variables in both formation are $n(n-1)/2+n$. However, $u_{ij}$ can be removed if the corresponding coefficient ($b_{ij}$ in \eqref{eq:omip} or $w_{ij}$ in \eqref{prob:mod2}) is 0. 
Since $w_{ij}$ is more likely to be zero than $b_{ij}$, we expect that the number of decision variables of the formulation for the modified problems \eqref{prob:mod2} is often much smaller than that of the original one. 
Although both programming formulations are NP-hard, our observation based on the simulation study in Section \ref{sec:synthetic} is that the reduction of the number of decision variables often lead to less computation in solving MIP. 

\section{Numerical Study on Synthetic Networks}\label{sec:synthetic}

This section compares three methods:
\begin{itemize}
    \item Original-MIP: the mixed integer programming formulation for the original D-optimality objective function in Section \ref{sec:mip}. 
    \item Modified-MIP: the mixed integer programming formulation 
    for the modified D-optimality objective function
    in Section \ref{sec:modified-mip}.
    \item Random: randomly allocate -1 or 1 with equal weights to each node.
\end{itemize}

We generate random networks to compare the three methods. 
For a network with $n$ nodes, the $n\times n$ adjacent matrix records the edges of this network. 
We randomly assign 0 and 1 to the upper or lower off-diagonal entries of this matrix. 
The proportion of ones is specified to be $p$. 
As defined in Section 2, the zero entry means that the two corresponding nodes are not adjacent, whereas one entry means the opposite. 
The proportion $p$ is referred to as the density of this network. 
Once we construct the designs using the three methods for a given network, we compare the designs on three aspects, computational efficiency, D-efficiency, and the empirical variance of the estimated $\beta$ in \eqref{eq:gcar}.

\begin{figure}
      \centering
      \includegraphics[width=0.99\textwidth]{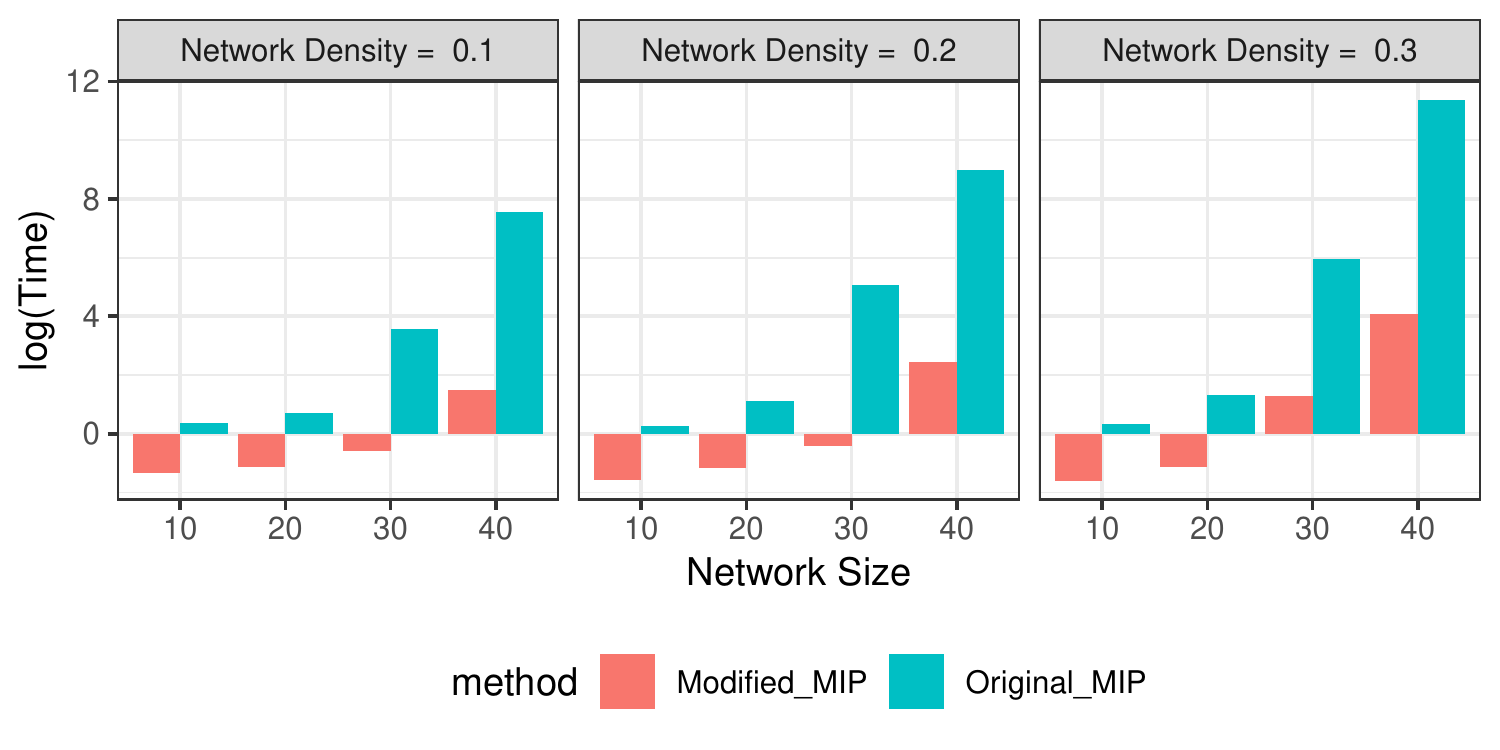}
      \captionof{figure}{Logarithm of running times in seconds of original-MIP and modified-MIP. For original-MIP, $\rho = 0.2$;
      for modified MIP, $\alpha = 0.6$. }
      \label{fig:density}
      \end{figure}

In terms of computational efficiency, we compare the computational time of solving the objective functions of original-MIP and modified-MIP with GUROBI solver (\url{http://www.gurobi.com/}). 
We set the network correlation coefficient to be $\rho=0.2$ for the original-MIP and the tuning parameter to be $\alpha=0.6$ for the modified-MIP. 
For both methods, the longest allowable running time is 24 hours. 
Exceeding that limit, the solver is terminated whether it reaches the optimal solution or not. 
As pointed out earlier, the number of decision variables of the modified-MIP is likely much smaller than that of original-MIP, thus the modified-MIP would take less time to run than original-MIP. 
To confirm this, the running time (in seconds) of these two methods with networks of different sizes are given in Figure \ref{fig:density}. 
It shows that the run time required to solve original-MIP increases dramatically as both network size and density increase. 
For the small networks in this simulation, the original-MIP is already so time-consuming thus it is not practical to be applied to the real-world social networks whose size is usually of thousands. 
Since the tuning parameter $\alpha$ determines the feasible region of the modified-MIP, we conduct an additional simulation whose results are shown in Table \ref{tab:alpha} in Appendix A.3. 
It implies that the value of $\alpha$ does not affect the computational time of modified-MIP significantly.

Next, we compare the three design methods using D-efficiency, whose value in \eqref{eq:d-eff} depends on the true correlation parameter $\rho$ of the CAR model.
Thus, we pick four different values for $\rho$ to 0, 0.1, 0.2, and 0.3, respectively.
We randomly generate a network of size $n=50$ and density $p=0.1$. 
For the modified-MIP, we vary the value of the tuning parameter $\alpha$ from 0.6, 0.7, to 0.8, and the running time of each case is less than one minute in GUROBI. 
For the random design, because $\mathrm{P}(x_i=1)=\mathrm{P}(x_i=-1)=1/2$, the expected D-efficiency is given by
\[
\frac{(1-\rho)(\sum^n_{i=1} m_i)^2-(1-\rho)^2\sum^n_{i=1} m^2_i}{(1-\rho)(\sum^n_{i=1}m_i)^2+(1-\rho)\rho\sum^n_{i=1}m_i\sum^n_{i=1}\sum^n_{j=1}w_{ij}}.,
\]
which only depends on the network. 
For the parameter $\rho$ in the original-MIP, we use $\rho=0.2$ (not necessarily equal to the true value of $\rho$ used in computing the D-efficiency) to the generate design. 
Since the original-MIP can be extremely time-consuming to obtain, we report the solution after 24 hours with an optimality gap of 10.0863\%. 
This optimality gap is defined as $(ub-lb)/ub$, where $ub$ and $lb$ are upper and lower bounds of the objective function. 
A smaller optimality gap indicates that the objective value corresponding to the current solution is closer to the true optimal objective value. 
Table \ref{tab:deff} gives the D-efficiency values.  
Both MIP based methods are better than the random design, especially when the correlation parameter is not zero. 
Also, the D-efficiency of modified-MIP under different $\alpha$ values is comparable with the original-MIP method.  

      \begin{table}[ht]
      \centering
      \caption{The D-Efficiency measures of different methods for  a network of size $n=50$ with density $p=0.1$}
      \label{tab:deff}
      \begin{tabular}{|c|cccc|}
      \hline Method  & $\rho = 0$ & $\rho = 0.1$ & $\rho = 0.2$ & $\rho = 0.3$ \\ 
    \hline Random & 0.98 & 0.90 & 0.82 & 0.76 \\ 

 Original-MIP&  1.00 & 0.96 & 0.93 & 0.90 \\
    Modified-MIP ($\alpha = 0.6$) & 1.00 & 0.96 & 0.93 & 0.90 \\  
    Modified-MIP ($\alpha = 0.7$)  & 1.00 & 0.96 & 0.92 & 0.90 \\ 
  Modified-MIP ($\alpha = 0.8$)  & 1.00 & 0.96 & 0.92 & 0.90 \\ 
      \hline 
      \end{tabular}
      \end{table}

Since the aim of A/B testing is to accurately estimate the treatment effect $\beta$ in \eqref{eq:gcar}, we assess the empirical variance of $\hat\beta$ to compare the three designs. 
Using the random network we investigated in Table \ref{tab:deff}, and a pre-specified correlation coefficient $\rho$, we generate $\pmb\delta$ from the multivariate normal distribution as expressed in \eqref{eq:mvn}. 
Set the values of the parameters as $\beta_0=0$, $\beta=2$ and $\sigma^2=1$ in the model \eqref{eq:gcar}. 
Given the design of $x_i$, we generate the response $y_i$ for each node as in \eqref{eq:gcar}.  
An estimate of $\beta$ can be obtained by fitting a linear regression model (LM) or the CAR model. 
For the CAR model, the parameters $\beta_0$, $\beta$ and $\rho$ are estimated by the maximum likelihood method using \verb|R| package \verb|spdep| \citep{bivand2005spdep}.
By repeating this procedure 500 times, we measure the accuracy of an estimate by calculating its sample variance :
\[
\hat{\mathrm V}(\hat\beta)=\frac{1}{499}\sum^{500}_{l=1}\left(\hat\beta_l-\frac{1}{500}\sum^{500}_{l=k}\hat\beta_k\right)^2,
\]
where $\hat\beta_1, \ldots, \hat\beta_{500}$ are the 500 copies of the estimates for $\beta$. 
Under our response generation scheme, the bias of the estimate $\hat\beta$ is tiny for all the methods. Hence, the mean squared error of $\hat\beta$ is dominated by the variance of $\hat\beta$, and the bias is negligible. So we ignore the bias and report empirical variance only.
Notice that the modified-MIP and the original-MIP generate deterministic designs for a given network, to the contrary of the random design. 
To make the three approaches comparable, 100 random designs are generated and we report the average value of $\hat{\mathrm V}(\hat\beta)$ over the 100 random designs. 
The empirical variance values of $\hat\beta$ for each method are given in Table \ref{tab:var50}. 
As in Table \ref{tab:deff}, we also vary the value of the tuning parameter $\alpha$ from 0.6, 0.7, to 0.8 for the modified-MIP, and fix $\rho=0.2$ for the original-MIP to generate design. 
The value of $\rho$ used in design generation is not equal to its real value for generating responses when $\rho=0$, 0.1, or 0.3 in Table \ref{tab:var50}.
For the random design, we consider the $\beta$ estimated from both CAR and LM, although the data are generated from the CAR model. 
We summarize our main observations from Table \ref{tab:var50} as follows:
\begin{itemize}
    \item For a non-zero correlation coefficient (i.e.,$\rho>0$), the variances of MIP based methods are smaller than the random design.
    \item The variances resulted from modified-MIP are comparable with those from original-MIP. 
\end{itemize}

      
      \begin{table}[ht]
      \centering
      \caption{Empirical variances of $\hat\beta$ based on designs generated from different methods for a network of size $n=50$ and density $p=0.1$}
      \label{tab:var50}
      \begin{tabular}{|l||c|c|c|c|}
      \hline Method & \textbf{$\rho = 0$} & \textbf{$\rho = 0.1$} & \textbf{$\rho = 0.2$} & \textbf{$\rho = 0.3$} \\ 
      \hline
      Original-MIP & 0.0046  & 0.0043 & 0.0041 & 0.0038 \\ 
      Modified-MIP ($\alpha = 0.6$) & 0.0044  & 0.0043 & 0.0040 & 0.0037 \\ 
       Modified-MIP ($\alpha = 0.7$) & 0.0049 & 0.0043 & 0.0043 & 0.0038 \\ 
      Modified-MIP ($\alpha = 0.8$) & 0.0047 & 0.0042 & 0.0042 & 0.0038 \\
      Random (CAR) & 0.0047  & 0.0047 & 0.0046 & 0.0046 \\ 
     Random (LM) & 0.0045  & 0.0045 & 0.0044 & 0.0045 \\ 
      \hline 
      \end{tabular}
      \end{table} 
      
In summary, the performance of modified-MIP is comparable with that of original-MIP for small networks. 
But it is much more practical because it takes less time to generate a design using GUROBI and does not require the correlation coefficient value $\rho$.
Even though it does involve the tuning parameter $\alpha$, its the performance is robust to the value of $\alpha$. 
So we choose $\alpha=0.6$ in remaining numerical examples and only compare the random design and modified-MIP for the real-world networks in Section \ref{sec:real}.

\section{Numerical Study on Real-world Networks}\label{sec:real}

This section studies two examples of real-world networks. 
Section \ref{sec:ego} considers examples of ego networks from Facebook. 
Section \ref{sec:cluster} considers examples of large networks containing multiple clusters. 

\subsection{Ego Networks}\label{sec:ego}
\begin{figure}
\centering
\includegraphics[scale=0.31]{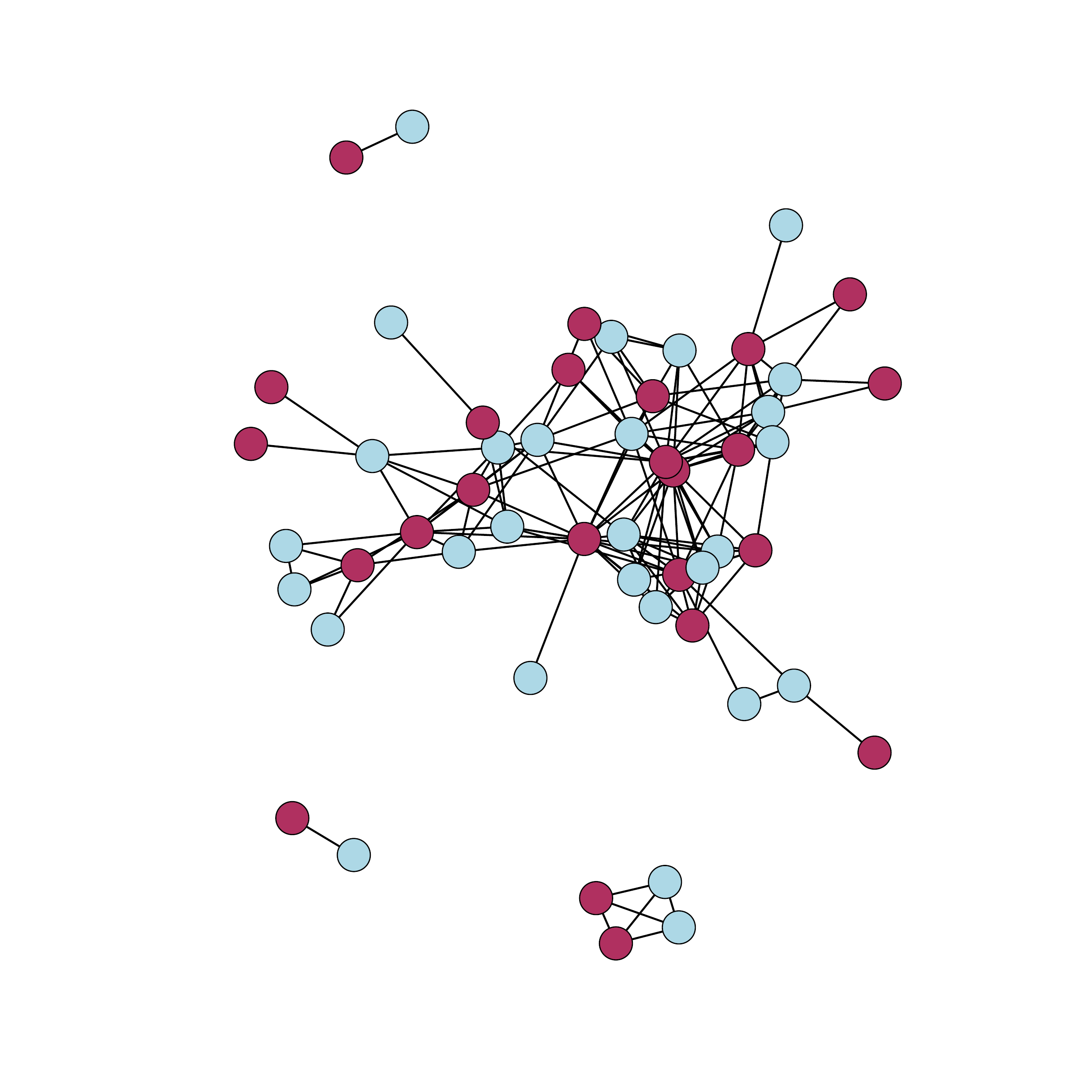}
\includegraphics[scale=0.31]{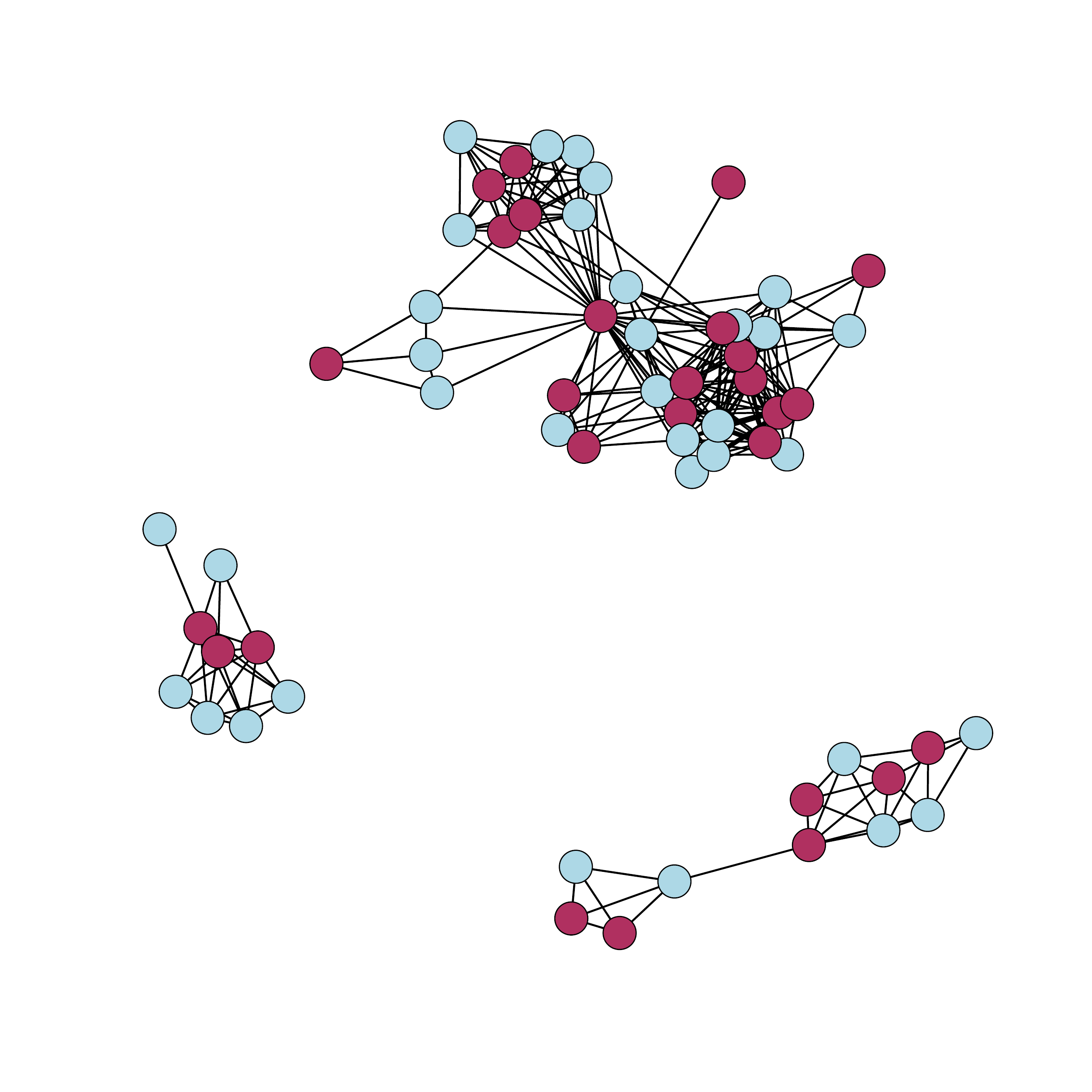}
\captionof{figure}{Left: ego network of size 52; right: ego network of size 61. The two different colors represents design allocation to -1 or 1 using modified-MIP.}
\label{fig:egonet}
\end{figure}

We consider five ego networks extracted from Facebook \citep{leskovec2012learning}.
These networks are collected from survey participants who used the Facebook App. 
Each ego network consists of a few focal nodes (i.e., egos) and the nodes that are directly connected to the egos. 
The sizes of these five ego networks are $n=52$, 61, 168, 333 and 224, respectively. 
The densities of them range from 0.05 to 0.15. 
We use the modified-MIP with $\alpha=0.6$ to construct the designs. 
As in Section \ref{sec:synthetic}, the maximum running time is set to be 24 hours in GUROBI. 
The optimality gaps of these five networks are 0\%, 0\%,  10.77\%, 36.87\%, and 39.45\%. 
Figure \ref{fig:egonet} depicts the designs generated from the modified-MIP for the ego networks of size 52 and 61. 

\begin{figure}
\centering
\includegraphics[width=0.99\textwidth]{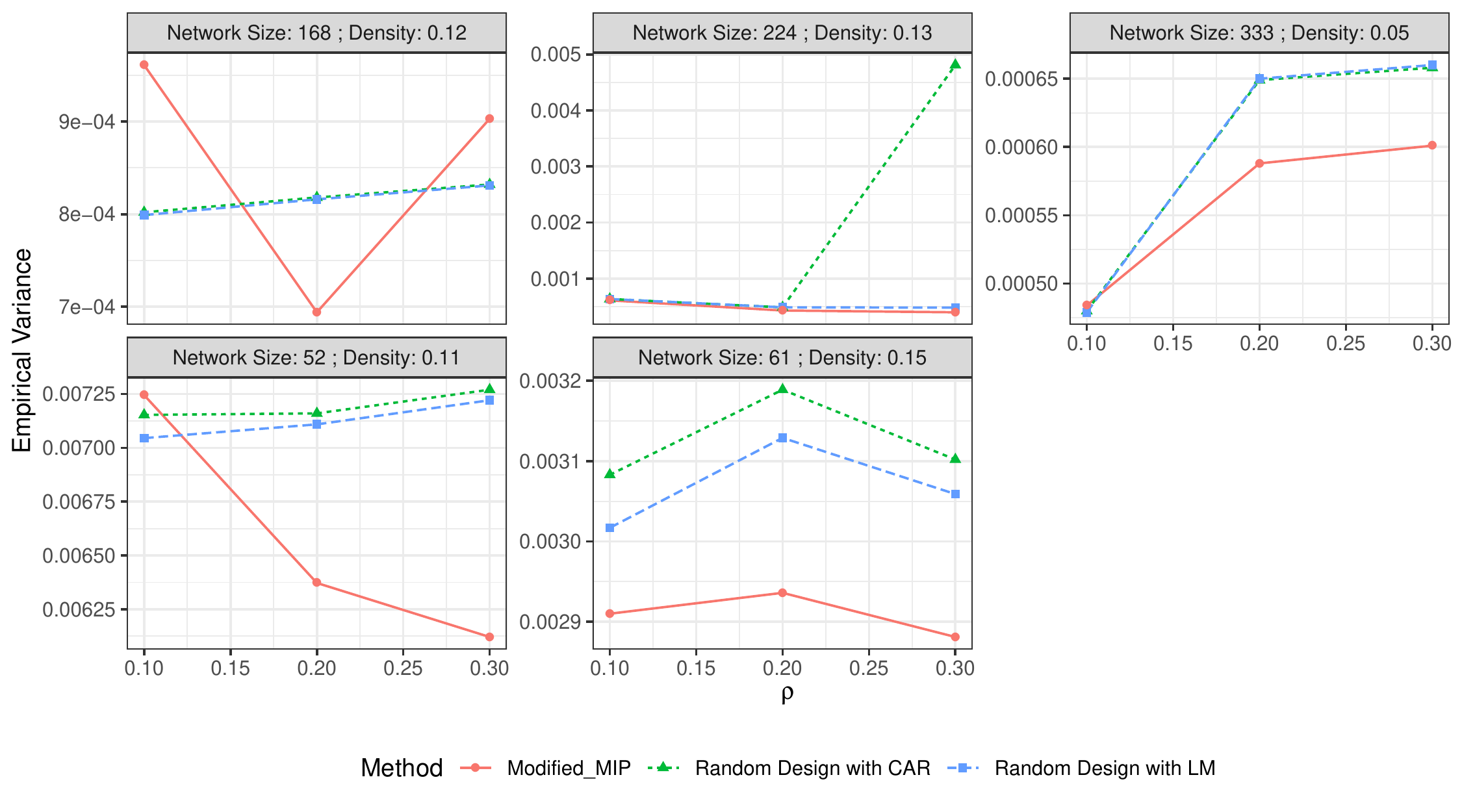}
\captionof{figure}{Empirical variances of $\hat{\beta}$ of five ego networks.}
\label{fig:var_comp}
\end{figure}

After obtaining designs for each ego network, we evaluate the empirical variances of $\hat\beta$, which is calculated the same as in Section \ref{sec:synthetic}. 
The true correlation parameter $\rho$ is specified to be 0.1, 0.2 or 0.3.
Three methods are compared: 1) the modified-MIP with $\hat\beta$ estimated using the CAR model; 2) the random design with $\hat\beta$ estimated using the CAR model; 3) the random design with $\hat\beta$ estimated using a linear regression model (LM).
The results are provided in Figure \ref{fig:var_comp}. We can see that in most cases the modified-MIP gives smaller variances than the random design when $\rho$ equals to 0.2 and 0.3. 
However, the advantage of modified-MIP is not significant when the correlation parameter $\rho$ is as small as $0.1$. 

\subsection{Large Networks with Disjoint Clusters} \label{sec:cluster}

We consider some sub-networks of the Facebook network available at \url{https://snap.stanford.edu/data/}. 
To sample the sub-networks, we randomly selected $K$ clusters, each cluster with approximately $50$ nodes.
The $K$ clusters form a network on which we apply our design. 
We pick $K=20$, $30$, and $40$ and the corresponding networks consist approximately 1000 to 2000 nodes. 
Then, we simulate the responses for each cluster of the large network. 
For each cluster, the correlation parameter $\rho$
is randomly generated from a uniform distribution $U(0.15, 0.25)$. Other parameters are specified the same as in Section \ref{sec:synthetic}.
The empirical variances are computed the same way as in Section \ref{sec:synthetic}. 
The estimate of $\beta$ can be obtained by fitting the CAR model or a linear regression model. 
Repeating this procedure 100 times, we assess the accuracy of the estimates by their sample variances. 
For the random design, we obtain the results from 100 random designs and record the average of the sample variances over these 100 random designs. 

We now discuss how to generate the MIP based designs for such large networks with disjoint clusters. 
For a network of size over 500, it is nearly impossible to directly obtain the MIP based designs. 
Since the large network is constructed by disjoint clusters, we can alternatively generate the MIP based design for each smaller cluster, and then combine them to obtain the design for the entire large network. 
Let $W_1, \ldots, W_K$ be the adjacent matrices of $K$ disjoint clusters. 
For each of them, we generate the two-level design using the modified-MIP described in Section 3.2.  
We denote the resulted design as $x_{ik}$ for $k=1,\ldots, K$ and $i=1, \ldots, n_k$, where $x_{ik}\in \{-1, 1\}$ is the treatment assignment for the $i$-th node from the $k$-th network.
Therefore, a combined design for the large network can be $\{c_k x_{ik}: i=1, \ldots, n_k, k=1, \ldots K\}$, where $c_k\in \{-1, 1\}$.
By varying the choices of $c_k$ for $k=1, \ldots, K$, the total number of combined designs for the large network is $2^K$. 
We take an additional step to choose the optimal value of $c_k$s by minimizing
\[
\mathrm{argmin}_{\bm c\in \{-1, 1\}^K }(\sum^K_{k=1}c_k\sum^{n_k}_{i=1}m_{ik}x_{ik})^2,
\]
where $\bm c=(c_1, \ldots, c_K)$, and $m_{ik}$ is the total number of neighbors for the $i$-th node from the $k$-th network, which leads that $\sum^{n_k}_{i=1}m_{ik}x_{ik}$ is a known constant.
The optimal solution of $\bm c$ gives the smallest value of $(\sum_i m_i x_i)^2$ in \eqref{eq:obj} for the large network. 
This optimization problem is the same as original-MIP in Section 3.1. For $K=20$, 30 or 40, this problem can be solved by GUROBI efficiently. 

\begin{figure}
\centering
\includegraphics[width=0.99\textwidth]{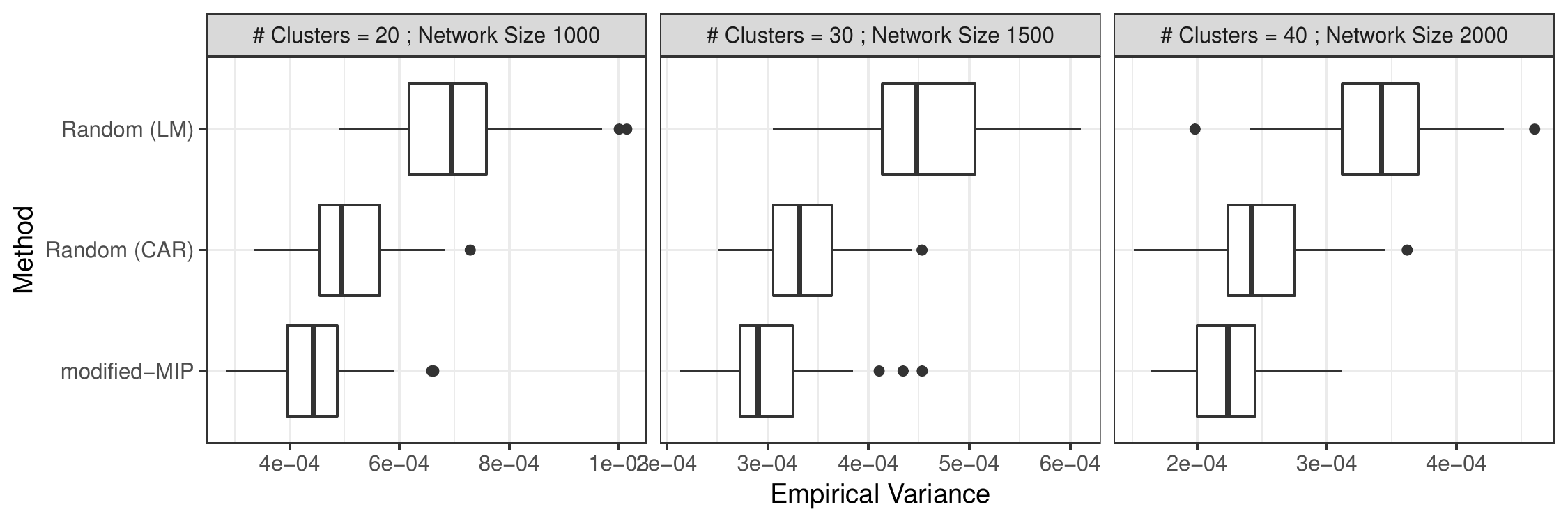}
\captionof{figure}{Boxplots of Empirical variances of $\hat{\beta}$ for large networks with clusters}
\label{fig:box}
\end{figure}

For $K=20$, 30 or 40, we generate 100 large networks with disjoint clusters as described above and compare the empirical variances of $\hat\beta$ of the three methods as in Section \ref{sec:ego}. The empirical variances of the 100 large networks for each $K$ are depicted in  Figure \ref{fig:box}. We see that the median of empirical variances of modified-MIP is the smallest among the three methods for all three cases. According to the results from the Wilcox's rank sum test, the median of modified-MIP is significantly smaller than that of the random design for each case. 




\section{Conclusion}

This paper proposes using the D-optimal design for A/B testing conducted on a social network. 
We use the CAR model to characterize the dependence of the responses from the adjacent nodes in a network. 
Mixed integer programming formulations are proposed to solve the D-optimal objectives and construct the designs. Numerical examples are provided to show the effectiveness of the proposed method. 
We plan to address the topics including how to apply the D-optimal design on the network for categorical responses, how to include the covariates information in the design on the network and how to construct sequential design on the network in the future. 

\bibliographystyle{asa}

\bibliography{projectREF.bib}

\newpage
\appendix
\section{Appendix}

\subsection{The derivation of the distribution $\pmb{\delta}$ \eqref{eq:mvn}}\label{sec:A1}
    According to Brook's Lemma, for any $\bm{Y} \in \mathbb{R}^n$, such that $\bm{Y} = (Y_1, Y_2, ... , Y_n)$ and any $\bm{Y}_0 \in \mathbb{R}^n$, such that $\bm{Y_0} = (Y_{01}, Y_{02}, ... , Y_{0n})$, where $Y_1, Y_2, ... , Y_n $ are random variables, and $Y_{01}, Y_{02}, ... , Y_{0n}$ is a copy of
    realizations of $Y_1, Y_2, ... , Y_n $, we have the following result: $$\frac{P(\bm{Y})}{P(\bm{Y}_0)} \propto \prod_{i=1}^{n} \frac{P(Y_i|Y_{01}, ..., Y_{0i-1}, Y_{i+1}, ..., Y_n)}{P(Y_{0i}|Y_{01}, ..., Y_{0i-1}, Y_{i+1}, ..., Y_n)}$$
      
    Let $\tilde{w}_{ij}=w_{ij}/m_i$ and $\sigma^2_i=\sigma^2/m_i$.  
    Under \eqref{eq:delta}, we have that
    \begin{align*}
    \frac{P(\bm{\delta})}{P(\bm{\delta}_0)} & \propto  \prod_{i=1}^{n} \frac{P(\delta_i|\delta_{01}, ..., \delta_{0i-1}, \delta_{i+1}, ..., \delta_n)}{P(\delta_{0i}|\delta_{01}, ..., \delta_{0i-1}, \delta_{i+1}, ..., \delta_n)} \propto  \prod_{i=1}^{n} \frac{\exp\{-\frac{1}{2\sigma_i^2}(\delta_i-\rho\sum_{j> i}\tilde{w}_{ij}\delta_j)^2\}}{\exp\{-\frac{1}{2\sigma_{0i}^2}(0-\rho\sum_{j> i}\tilde{w}_{ij}\delta_j)^2\}}\\ & \propto \exp\{ \sum_{i=1}^{n}[-\frac{1}{2\sigma_i^2}(\delta_i-\rho\sum_{j> i}\tilde{w}_{ij}\delta_j)^2 + \frac{1} {2\sigma_i^2} (\rho\sum_{j> i}\tilde{w}_{ij}\delta_j)^2] \} \\ & \propto \exp\{ \sum_{i=1}^{n}[-\frac{1}{2\sigma_i^2}(\delta_i^2-2\rho\delta_i\sum_{j> i}\tilde{w}_{ij}\delta_j + (\rho\sum_{j> i}\tilde{w}_{ij}\delta_j)^2 -(\rho\sum_{j> i}\tilde{w}_{ij}\delta_j)^2)] \} \\ & \propto \exp\{ \sum_{i=1}^{n}[-\frac{1}{2\sigma_i^2}(\delta_i^2-2\rho\delta_i\sum_{j> i}\tilde{w}_{ij}\delta_j)] \} \propto \exp\{ -\frac{1}{2\sigma^2} \sum_{i=1}^{n} m_i (\delta_i^2-2\rho\delta_i\sum_{j> i}\tilde{w}_{ij}\delta_j) \} \\ & \propto \exp\{ -\frac{1}{2\sigma^2}  (\sum_{i=1}^{n} m_i\delta_i^2-2\rho\sum_{i=1}^{n}\sum_{j> i}\delta_im_i\tilde{w}_{ij}\delta_j)\} \\ & \propto \exp\{ -\frac{1}{2\sigma^2}  (\sum_{i=1}^{n} m_i\delta_i^2-2\rho\sum_{i=1}^{n}\sum_{j> i}\delta_i\frac{m_i{w}_{ij}}{m_i}\delta_j) \} \\ & \propto \exp\{ -\frac{1}{2\sigma^2}  (\sum_{i=1}^{n} m_i\delta_i^2-2\rho\sum_{i=1}^{n}\sum_{j> i}\delta_iw_{ij}\delta_j) \} \propto \exp\{ -\frac{1}{2\sigma^2} \bm{\delta}^T (D-\rho W) \bm{\delta} \} 
    \end{align*}  
    Therefore, 
\[
\boldsymbol{\delta}=(\delta_1, \ldots, \delta_{n})^\top\sim \mathcal{MVN}_n(0, \sigma^2(D-\rho W)^{-1})
\]

\subsection{Proof of Proposition \ref{prop:dopt}}

Following the expression in \eqref{eq:dopt}, we have that

\begin{align*}
    &\mathrm{max}_{\bm x \in \{-1,1\}^n}[D(\bm x)] \\
    & = \mathrm{max}_{\bm x \in \{-1,1\}^n}[(1-\rho)\sum\limits_i m_i (\sum\limits_i m_i x_i^2-\rho \sum\limits_i \sum\limits_j  w_{ij} x_i x_j) - (1-\rho)^2(\sum\limits_i m_i x_i)^2]\\
    & = \mathrm{max}_{\bm x \in \{-1,1\}^n}[\sum\limits_i m_i \sum\limits_i m_i - \rho \sum\limits_i m_i \sum\limits_i \sum\limits_j w_{ij} x_i x_j - (1-\rho)(\sum\limits_i m_i x_i)^2],
    \end{align*}
    which is equivalent to 
    \begin{align*}
     & = \mathrm{min}_{\bm x \in \{-1,1\}^n}\left[\rho \sum\limits_i m_i\sum\limits_i \sum\limits_j  w_{ij} x_i x_j + (1-\rho)(\sum\limits_i m_i x_i)^2\right]\\
    & = \mathrm{min}_{\bm x \in \{-1,1\}^n}\left[\frac{\rho}{1-\rho} \sum\limits_i m_i\sum\limits_i \sum\limits_j  w_{ij} x_i x_j + (\sum\limits_i m_i x_i)^2\right]\\
    & = \mathrm{min}_{\bm x \in \{-1,1\}^n}\left[a \sum\limits_i \sum\limits_j w_{ij} x_i x_j + (\sum\limits_i m_i x_i)^2\right],
    \end{align*}
where $\bm x=(x_1, \ldots, x_n)$, 
    $x_i\in \{-1,1\}$, and $a = \frac{\rho}{1-\rho} \sum\limits_i m_i$.     

\subsection{Additional numerical results}\label{sec:A2}
Table \ref{tab:alpha} gives the time (in seconds) took for GUROBI to solve modified MIP for different network sizes and different choices of $\alpha$. Similar to the original-MIP, the running times of modified-MIP increase for larger networks. 

      \begin{table}[ht]
      \centering
      \caption{\footnotesize Running times (in sec) of modified-MIP for different values of $\alpha$ averaged across 20 network sizes $n$ keeping network density $p = 0.1$.}
      \label{tab:alpha}
      \begin{tabular}{|c||c|c|c|c|}
      \hline $\alpha$  & $n=20$ & $n=30$ & $n=40$ & $n=50$ \\ 
      \hline 0.55 & 0.47 & 1.63 & 3.61 & 13.26\\
      \hline 0.6 & 0.58 & 1.31  & 3.95 & 24.80 \\ 
      \hline 0.7 & 0.57 & 1.40  & 3.67 & 13.65 \\ 
      \hline 0.8 & 0.60 & 1.43 & 2.12 & 21.65 \\
      \hline 0.9 & 0.58 & 1.47  & 4.36 & 17.08 \\
      \hline 
      \end{tabular}
      \end{table} 
      


\end{document}